\documentclass[12pt,thmsa]{article}
\usepackage{sw20lart}

\input{tcilatex}
\begin{document}

\title{Single-particle Bell-type Inequality }
\author{A. Shafiee\thanks{%
E-mail: shafiee@theory.ipm.ac.ir} \ \ \ M. Golshani\thanks{%
E-mail: golshani@ihcs.ac.ir} \and Institute for studies in Theoretical
Physics and Mathematics \and ( P. O. Box 19395-5531, Tehran, Iran )}
\maketitle

\begin{abstract}
It is generally believed that Bell's inequality holds for the case of
entangled states, including two correlated particles or special states of a
single particle. Here, we derive a single-particle Bell's inequality for two
correlated spin states at two successive times, appealing to the statistical
independence condition in an ideal experiment, for a locally causal hidden
variables theory. We show that regardless of the locality assumption, the
inequality can be violated by some quantum predictions.
\end{abstract}

\section{\protect\smallskip Introduction}

After Bell derived his well- known inequality for a Bohmian version [1] of
EPR [2] ( hereafter called EPRB) thought experiment and showed its
inconsistency with quantum mechanics [3], most authors considered local
realism to be untenable, and attributed this inconsistency to the
non-locality present in nature. The entangled states, in these experiments,
are assumed to play a crucial role in the derivation of Bell's inequality.

In recent years, certain generalizations of Bell's inequality has been
proposed in which locality is supposed to be violated [4]. Some people, e.g.
Elitzur and Vaidman, have tried to prove non-locality without any appeal to
any inequality [5], and Hardy has extended this idea to the case of single
particles [6].

Although most of the works done on the single-particle case have been in the
direction of denying locality, there has been some attempts in the opposite
direction too. Works of Leggett and Garg [7], as well as Home and Sengupta
[8] are of this category. The former authors assume locality, but challenge
the applicability of quantum mechanics to the macroscopic phenomena. The
latter try to show that Bell's inequality is derivable as a general
consequence of non-contextual hidden variables theories. To show this, they
have considered an entangled wavefunction which is a superposition of two
factorized states in the general form of $\Psi =\stackrel{2}{\stackunder{i=1%
}{\sum }}c_{i}u_{i}v_{i}$ , where the $u_{i}$ and the $v_{i}$ are
eigenstates of the orbital and spin angular momentum, respectively, of a
single valence electron. It is claimed that it is possible to drive Bell's
inequality for every entangled state and in this sense, there exists a
particular way of preparing single particle states [9].

In our proposed experiment, however, we consider a source of microscopic
spin $1/2$ particles for which the quantum state can be expressed as a sum
of two individual spin states and is changed at two successive times. Then,
we derive Bell's inequality as a consequence of the statistical independence
condition for the ideal joint probability functions of a locally causal
hidden variables theory. The meaning of this condition will be made explicit
in the following section.

\section{Argument}

Let us consider a primary source which emits spin $1/2$ particles that are
polarized along the x-axis, i.e., $\mid \Psi _{0}\rangle =\frac{1}{\sqrt{2}}%
\left[ \mid z+\rangle +\mid z-\rangle \right] $ where $\mid z+\rangle $ and $%
\mid z-\rangle $ are the two base vectors which correspond to the two
eigenvectors of $\sigma _{z}$. There is a relatively large time interval
between the emission of successive particles. Thus, we assume that only one
particle passes in sequence through two analyzers (Stern-Gerlach
apparatuses) \textbf{M}$_{1}(\widehat{a})$ along the angle $\widehat{a}$ at t%
$_{1}$ and \textbf{M}$_{2}(\widehat{b})$ or \textbf{M}$_{2}^{^{\prime }}(%
\widehat{b})$ along the angle $\widehat{b}$ at t$_{2}$ (t$_{2}>$ t$_{1}$),
relative to the z-axis. Then, the particle coming out of \textbf{M}$_{2}(%
\widehat{b})$ or \textbf{M}$_{2}^{^{\prime }}(\widehat{b})$, is detected by
one of the detectors D$_{++}$, D$_{+-}$, D$_{-+}$ or D$_{--}$ (at a time
larger than t$_{2}$) and we see one of these detectors to be flashing
(Fig.1). One can assign a value $A$ and a value $B$, respectively, to the
spin components of the particle along $\widehat{a}$ at t$_{1}$ and $\widehat{%
b}$ at t$_{2}$ ($A,B=\pm 1$, in units of $\hbar /2$), when one of the
detectors flashes. A flash in D$_{++}$, e.g., means that $A=+1$ and $B=+1$.
No component is filtered or blocked at t$_{1}$and t$_{2}$, and the result
becomes known only at a time after t$_{2}$. From the detector that flashes,
one can infer the values of $A$ and $B$ which correspond with the spin up ($%
+1$) or down ($-1$) of the particle at t$_{1}$ and t$_{2}$, respectively. We
assume that the spin vector is a constant of motion in all stages of our
experiment.

In quantum mechanics, we represent the physical states (spin states) of each
particle at t$_{1}$ and t$_{2}$ by $\mid \varphi _{A}^{(t_{1})}\rangle $ and 
$\mid \varphi _{B}^{(t_{2})}\rangle $, respectively. These individual spin
states are defined as

\[
\mid \varphi _{+}^{(t_{i})}\rangle =\cos \frac{\widehat{\theta }_{i}}{2}\mid
z+\rangle +\sin \frac{\widehat{\theta }_{i}}{2}\mid z-\rangle 
\]
and

\[
\mid \varphi _{-}^{(t_{i})}\rangle =-\sin \frac{\widehat{\theta }_{i}}{2}%
\mid z+\rangle +\cos \frac{\widehat{\theta }_{i}}{2}\mid z-\rangle 
\]
where $i=1,2\ ;\widehat{\theta }_{1}=\widehat{a}$ and $\widehat{\theta }_{2}=%
\widehat{b}$.

In an ideal experiment, the probability that we have the value $A$ at t$_{1}$
and the value $B$ at t$_{2}$, as a result of the joint analysis of the spin
components $\sigma _{a}^{(t_{1})}=\overrightarrow{\sigma }^{(t_{1})}.%
\widehat{a}$ and $\sigma _{b}^{(t_{2})}=\overrightarrow{\sigma }^{(t_{2})}.%
\widehat{b}$, respectively, is

\begin{equation}
P^{(t_{1},t_{2})}(\sigma _{a}^{(t_{1})}=A,\sigma _{b}^{(t_{2})}=B|\widehat{a}%
,\widehat{b},\Psi _{0})=|\langle \Psi _{0}|\varphi _{A}^{(t_{1})}\rangle
|^{2}|\langle \varphi _{A}^{(t_{1})}|\varphi _{B}^{(t_{2})}\rangle |^{2} 
\tag{1}
\end{equation}

This probability depends on the state preparation of the source (denoted by $%
\Psi _{0}$) and the orientation of the Stern-Gerlach (SG) apparatuses at t$%
_{1}$ and t$_{2}$. It will be abbreviated as $P^{(t_{1},t_{2})}(A,B|\widehat{%
a},\widehat{b},\Psi _{0})$, and can be derived to be

\begin{equation}
P^{(t_{1},t_{2})}(A,B|\widehat{a},\widehat{b},\Psi _{0})=\frac{1}{4}\left(
1+A\sin \widehat{a}\right) \left[ 1+AB\cos \left( \widehat{a}-\widehat{b}%
\right) \right]  \tag{2}
\end{equation}

One can obtain the probabilities for the values at t$_{1}$ or t$_{2}$ by
summing both sides of (2) over appropriate parameters. In this way, the
probability of having the value $\sigma _{a}^{(t_{1})}=A$ at t$_{1},$ is

\begin{equation}
P^{(t_{1})}(A|\widehat{a},\Psi _{0})=\stackunder{B=\pm 1}{\sum }%
P^{(t_{1},t_{2})}(A,B|\widehat{a},\widehat{b},\Psi _{0})=\frac{1}{2}\left(
1+A\sin \widehat{a}\right)  \tag{3}
\end{equation}

Regardless of the result at t$_{1}$, the probability of having the value $%
\sigma _{b}^{(t_{2})}=B$ at t$_{2}$, is

\begin{eqnarray}
P^{(t_{2})}(B|\widehat{a},\widehat{b},\Psi _{0}) &=&\stackunder{A=\pm 1}{%
\sum }P^{(t_{1},t_{2})}(A,B|\widehat{a},\widehat{b},\Psi _{0})  \nonumber \\
&=&\frac{1}{2}\left[ 1+B\sin \widehat{a}\cos \left( \widehat{a}-\widehat{b}%
\right) \right]  \tag{4}
\end{eqnarray}

It is also obvious from the relations (2) and (3) that the conditional
probability of the value $B$ at t$_{2}$ is equal to

\begin{equation}
P^{(t_{2})}(B|\widehat{a},\widehat{b},A,\Psi _{0})=\frac{1}{2}\left[
1+AB\cos \left( \widehat{a}-\widehat{b}\right) \right]  \tag{5}
\end{equation}

After the particle came out of the first SG apparatus, its spin state
changes to a new state $\varphi _{A}^{(t_{1})}$. Since no result is detected
at t$_{1}$, it is not obvious at this step that the new state is $\varphi
_{+}^{(t_{1})}$ or $\varphi _{-}^{(t_{1})}$. However, we can generally
interpret the relation (5) in terms of the new preparation as

\begin{equation}
P^{(t_{2})}(B|\widehat{a},\widehat{b},A,\Psi _{0})=P^{(t_{2})}(B|\widehat{b}%
,\varphi _{A}^{(t_{1})})  \tag{6}
\end{equation}

The probability distributions (4) and (5) which are defined at t$_{2}$,
depend on the orientation of the SG apparatus at t$_{1}.$ This can be
explained by the correlation which exists between the statistical values of
the spin components of the particle at two successive times t$_{1}$ and t$%
_{2}$ [10]. This correlation is due to the non-factorized form of the joint
probability (1) and it originates from a new state preparation of the
particle at t$_{1},$ which is denoted by $\varphi _{A}^{(t_{1})}$ in the
relation (6). There is no room for non-locality in this experiment, because
the events at t$_{1}$ and t$_{2}$ are time-like separated, and when the
particle is coming out from \textbf{M}$_{1}(\widehat{a})$ at t$_{1},$ there
is no particle at t$_{2},$ and the communication of information from t$_{1}$
to t$_{2}$ is done by the particle itself. Thus, the problem of non-locality
does not arise.

In an actual experiment and for a massive spin $1/2$ particle (like an
electron), if we assume that the SG apparatuses are sufficiently efficient,
the probability of detecting a result by one of the detectors D$_{AB}$ ($%
A=\pm 1$ and $B=\pm 1$) can be given by the following relation

\begin{equation}
P_{\exp }(D_{AB})=\eta _{D}FP^{(t_{1},t_{2})}(A,B|\widehat{a},\widehat{b}%
,\Psi _{0})  \tag{7}
\end{equation}
where, $\eta _{D}$ is the efficiency of the detector D$_{AB}$ (the
efficiencies of all the detectors are assumed to be the same), and $F$ is
the overall probability that a single particle emitted by the source will
enter one of the detectors D$_{AB}$.

This probability is equal to

\begin{equation}
F=f_{1}f_{21}f_{D2}  \tag{8}
\end{equation}
where, $f_{1}$ is the probability that a particle will enter \textbf{M}$_{1}(%
\widehat{a})$ at t$_{1}$; $f_{21}$ is the conditional probability that the
particle will enter \textbf{M}$_{2}(\widehat{b})$ or \textbf{M}$%
_{2}^{^{\prime }}(\widehat{b})$ at t$_{2}$, after it has passed through 
\textbf{M}$_{1}(\widehat{a})$ at t$_{1}$; and $f_{D2}$ is the conditional
probability that the particle will reach a detector, when it has already
passed through \textbf{M}$_{2}(\widehat{b})$ or \textbf{M}$_{2}^{^{\prime }}(%
\widehat{b})$ at t$_{2}$. These functions are, in fact, the collimator
efficiencies and are proportional to the collimator acceptance solid angles.
A detection is present, when all the functions $f_{1}$, $f_{21}$, and $%
f_{D2} $ are different from zero.

Here, we have not introduced a correlation factor within the probability of
detection $P_{\exp }(D_{AB})$, because unlike the case of two particles
involved in the regular Bell-type experiments, the initial state is not an
entangled one. There can be a complete correlation, however, between the
values of the spin components in the same directions at t$_{1}$ and t$_{2}$,
because the spin of the particle is assumed to be conserved.

Now, we consider a locally causal hidden variables theory, as used by Bell
and others [11]. In this context, we assume that the spin state of particle
is described by a function of a collection of hidden variabless called $%
\lambda ,$ which belongs to a space $\Lambda .$ The parameter $\lambda $
contains all the information which is necessary to specify the spin state of
the system. Using the spin state of the particle, it would be possible to
define the probability measures and the corresponding mean values on $%
\Lambda .$ In this way, we can define the mean value of the product of the
values of the spin components for the particle at times t$_{1}$ and t$_{2}$
along $\widehat{a}$ and $\widehat{b}$, respectively, as

\begin{equation}
E^{(t_{1},t_{2})}(\widehat{a},\widehat{b},\lambda )=\stackunder{A,B=\pm 1}{%
\sum }AB\ \wp ^{(t_{1},t_{2})}(A,B|\widehat{a},\widehat{b},\lambda )  \tag{9}
\end{equation}
where, $\wp ^{(t_{1},t_{2})}(A,B|\widehat{a},\widehat{b},\lambda )$ is the
joint probability of the values $A$ and $B,$ corresponding to the spin
components of the particle along $\widehat{a}$ at t$_{1}$ and $\widehat{b}$
at t$_{2}$, respectively. As a consequence of the principles of the
probability theory, the joint probability $\wp ^{(t_{1},t_{2})}(A,B|\widehat{%
a},\widehat{b},\lambda )$ is equivalent to the following product form

\begin{equation}
\wp ^{(t_{1},t_{2})}(A,B|\widehat{a},\widehat{b},\lambda )=\wp ^{(t_{1})}(A|%
\widehat{a},\lambda )\wp ^{(t_{2})}(B|\widehat{a},\widehat{b},A,\lambda ) 
\tag{10}
\end{equation}

Now, we define the statistical independence condition, for an ideal case at
the hidden variables level, as the conjunction of the following two
assumptions:

\begin{quote}
\underline{$\mathbf{C}_{1}$}. For definite settings of the two SG
apparatuses at t$_{1}$ and t$_{2}$, the probability of having a value $B$ at
t$_{2}$ is independent of the value $A$ at t$_{1}.$

\underline{$\mathbf{C}_{2}$}. The probability of having a value $B$ at t$%
_{2} $ is independent of the setting of the SG apparatus at t$_{1}.$
\end{quote}

The assumptions $\mathbf{C}_{1}$ and $\mathbf{C}_{2}$ will be structurally
the same as the outcome independence and parameter independence,
respectively, in Shimony's terminology for a two-particle Bell state [12],
if in their definitions the times t$_{1}$ and t$_{2}$ are replaced by two
spatially separated locations $L_{1}$ and $L_{2}$ where the values $A$ and $%
B $ are respectively assigned to the spin components of particle $1$ along $%
\widehat{a}$ and particle $2$ along $\widehat{b}$. As is the case for a
two-particle entangled state, the conjunction of these two assumptions leads
to the factorization of the corresponding joint probability. Consequently,
the joint probability (10) takes following form

\begin{equation}
\wp ^{(t_{1},t_{2})}(A,B|\widehat{a},\widehat{b},\lambda )=\wp ^{(t_{1})}(A|%
\widehat{a},\lambda )\wp ^{(t_{2})}(B|\widehat{b},\lambda )  \tag{11}
\end{equation}

Both the assumptions $\mathbf{C}_{1}$ and $\mathbf{C}_{2}$ are violated by
quantum mechanics, as is obvious from the relations (4) and (5). The
negation of $\mathbf{C}_{1}$ and $\mathbf{C}_{2}$ at the quantum level is
caused by the statistical dependence of the probability functions at t$_{2}$
on the condition(s) generated as a result of the preparation made for the
spin state of particle at t$_{1}.$

We assume, however, that the hidden probabilities have a classical character
[13]. That is, for a definite system, there exist hidden statistical
distributions for the values of the spin components of a particle along
definite directions which depend \textit{only} on the initial state of the
system (represented by some hidden variables) and do not depend on any
preparation procedure before the measurement.

As an example, one can suppose that the spin state of a particle depends on
its path and has certain projections, e.g., along $\widehat{a}$ at t$_{1}$
and $\widehat{b}$ at t$_{2}$ which we call, respectively, $s(\overrightarrow{%
x}(t_{1}),\widehat{a})$ and $s(\overrightarrow{x}(t_{2}),\widehat{b})$.
Here, the hidden variables (denoted by $\lambda $) are the initial position
coordinates $\overrightarrow{x}(0)$ by which $\overrightarrow{x}(t)$ can be
determined at any arbitrary time. Now, one can build a statistics for the
values of the spin components $s(\overrightarrow{x}(t_{1}),\widehat{a})$ at t%
$_{1}$ and $s(\overrightarrow{x}(t_{2}),\widehat{b})$ at t$_{2}$, based on
the different possible initial positions $\overrightarrow{x}(0)$. This
representative example shows what we really mean by introducing the hidden
probabilities $\wp ^{(t_{1})}(A|\widehat{a},\lambda )$ and $\wp ^{(t_{2})}(B|%
\widehat{b},\lambda )$.

Accordingly, the validity of the assumptions $\mathbf{C}_{1}$ and $\mathbf{C}%
_{2}$ in a locally causal hidden variables theory is a consequence of the
fact that any information about the values of the spin components of
particle at t$_{1}$ and t$_{2}$ originates from $\lambda $ and that the
first apparatus \textbf{M}$_{1}(\widehat{a})$ does not affect the \textit{%
spin state} of the particle. This means that the past history of the
particle is based on $\lambda $ alone, and the spin values as well as the
setting of the SG apparatus at t$_{1}$, have no role in specifying the spin
state of particle at t$_{2}$. By averaging over $\lambda $, however, a new
description may be needed in which the state preparation of the system at
any time plays an important role in the description of the state of the
system. Our argument shows that it cannot be expected \textit{a priori} that
the same situation holds at a sub-quantum level, too.

Inserting (11) into (9), one gets

\begin{equation}
E^{(t_{1},t_{2})}(\widehat{a},\widehat{b},\lambda )=E^{(t_{1})}(\widehat{a}%
,\lambda )\ E^{(t_{2})}(\widehat{b},\lambda )  \tag{12}
\end{equation}
where

\[
E^{(t_{1})}(\widehat{a},\lambda )=\stackunder{A=\pm 1}{\sum }A\wp
^{(t_{1})}(A|\widehat{a},\lambda ) 
\]
and

\[
E^{(t_{2})}(\widehat{b},\lambda )=\stackunder{B=\pm 1}{\sum }B\wp
^{(t_{2})}(B|\widehat{b},\lambda ) 
\]

Here, $E^{(t_{1})}(\widehat{a},\lambda )$ and $E^{(t_{2})}(\widehat{b}%
,\lambda )$ are, respectively, the mean values of the spin components of
particle along $\widehat{a}$ at t$_{1},$ and $\widehat{b}$ at t$_{2}$.

In an actual experiment, one can define the following correspondence relation

\begin{equation}
P_{\exp }(D_{AB})=\int_{\Lambda }\wp _{\exp }(D_{AB},\lambda )\rho (\lambda
)d\lambda  \tag{13}
\end{equation}
where the probability density $\rho (\lambda )$ is defined over the space $%
\Lambda $ ($\int_{\Lambda }\rho (\lambda )d\lambda =1$) and $\wp _{\exp
}(D_{AB},\lambda )$ is defined by

\begin{equation}
\wp _{\exp }(D_{AB},\lambda )=\eta _{D}F\wp ^{(t_{1},t_{2})}(A,B|\widehat{a},%
\widehat{b},\lambda )  \tag{14}
\end{equation}

This relation shows that regardless of what is assumed in the context of the
hidden variables theory, the events in the future may depend on what was
occurred in the past. This situation happens, when at least one of the
conditional probabilities $f_{21}$ or $f_{D2}$ is not equal to one; rather
it is less than one. Thus, in an actual experiment, the statistical
independence condition cannot be satisfied, in general, even if the
assumptions $\mathbf{C}_{1}$ and $\mathbf{C}_{2}$ hold at the hidden
variabless level. Since $F$ is independent of the content of the theory
defining $\lambda $ and is only determined experimentally, this possibility
remains open to use the relation (11) for our next purposes.

In an ideal experiment, if we consider the statistical independence
condition at the hidden variables level (the relation (11)), it is generally
possible to reproduce the quantum mechanical predictions (see appendix).
Then, for the definite settings of the SG apparatuses along $\widehat{a}$ or 
$\widehat{a^{\prime }}$ at t$_{1}$ and $\widehat{b}$ or $\widehat{b^{\prime }%
}$ at t$_{2},$ one can obtain Bell's inequality- in Shimony's way of
deriving [12]- in the following form

\begin{equation}
\mid E^{(t_{1},t_{2})}(\widehat{a},\widehat{b},\lambda )+E^{(t_{1},t_{2})}(%
\widehat{a},\widehat{b^{\prime }},\lambda )+E^{(t_{1},t_{2})}(\widehat{%
a^{\prime }},\widehat{b^{\prime }},\lambda )-E^{(t_{1},t_{2})}(\widehat{%
a^{\prime }},\widehat{b},\lambda )\mid \leq 2  \tag{15}
\end{equation}

Multiplying (15) through the probability density $\rho (\lambda )$ and
integrating over $\Lambda $, we get the following inequality at the quantum
level

\begin{equation}
\mid \langle \sigma _{a}^{(t_{1})}\sigma _{b}^{(t_{2})}\rangle +\langle
\sigma _{a}^{(t_{1})}\sigma _{b^{\prime }}^{(t_{2})}\rangle +\langle \sigma
_{a^{\prime }}^{(t_{1})}\sigma _{b^{\prime }}^{(t_{2})}\rangle -\langle
\sigma _{a^{\prime }}^{(t_{1})}\sigma _{b}^{(t_{2})}\rangle \mid \leq 2 
\tag{16}
\end{equation}
where, e.g., we have set the quantum expectation values as

\begin{equation}
\langle \sigma _{a}^{(t_{1})}\sigma _{b}^{(t_{2})}\rangle =\int_{\Lambda
}E^{(t_{1},t_{2})}(\widehat{a},\widehat{b},\lambda )\ \rho (\lambda )\
d\lambda  \tag{17}
\end{equation}

Using the definition of $\langle \sigma _{a}^{(t_{1})}\sigma
_{b}^{(t_{2})}\rangle _{\exp }$ for an actual experiment, as

\[
\langle \sigma _{a}^{(t_{1})}\sigma _{b}^{(t_{2})}\rangle _{\exp }=%
\stackunder{A,B=\pm 1}{\sum }AB\ P_{\exp }(D_{AB}) 
\]
and the relations (2) and (7), one gets

\begin{equation}
\langle \sigma _{a}^{(t_{1})}\sigma _{b}^{(t_{2})}\rangle _{\exp }=\eta
_{D}F\cos \left( \widehat{a}-\widehat{b}\right)  \tag{18}
\end{equation}

Other experimental expectation values are similarly obtained. Now, if we
choose all angles $\widehat{a},\widehat{a^{\prime }},\widehat{b}$ and $%
\widehat{b^{\prime }}$ in the xz-plane and let $|\widehat{a}-\widehat{b}|=|%
\widehat{a}-\widehat{b^{\prime }}|=|\widehat{a^{\prime }}-\widehat{b^{\prime
}}|=\alpha ,\ $and $|\widehat{a^{\prime }}-\widehat{b}|=3\alpha $, then, for
an actual experiment, (16) reduces to

\[
\eta _{D}F\mid 3\cos \alpha -\cos 3\alpha \mid \leq 2 
\]

This can be violated, if $\eta _{D}F>\frac{1}{\sqrt{2}}$. If we assume that $%
\eta _{D}\simeq 1$ (which could be achived in actual experiments), this
means that the overall probability of detection should be greater than 71\%.
Thus, under these conditions, the factorizability relation (11) for a
locally causal hidden variables theory leads to inconsistency with quantum
predictions. This shows that one cannot base the past history of the system
on $\lambda $ alone, and it is possible that the concept of state
preparation is an intrinsic property of microscopic states.

\section{Conclusion}

As was indicated by Shr$\stackrel{..}{\text{o}}$dinger [14], \textit{the
quantum entanglement is the characteristic trait of quantum mechanics.}
Emphasizing the significance of the quantum entanglement, Shimony argued
that outcome independence is violated for a two-particle singlet state [12].
The incompatibility of the quantum mechanical predictions with the local
realistic hidden variables theories has been frequently reported for one
[15], two [16] and more than two-particle \textit{entangled} states [17, 18].

The entangled states lead to the correlation between different eigenvalues
corresponding to the factorized eigenstates, but, it is important to notice
that the existence of correlation is not limited to the entangled states.
Here, we have shown another possibility. The correlation between the
statistical values of the spin components of a single particle at two
successive times can be related to the statistical dependence of the
probability distributions on the earlier preparation. In this sense, there
is a point of similarity between all the experiments concerning Bell's
inequality, if one uses the state preparation point of view. The difference
appears when we distinguish what kind of state preparation is the source of
correlation. For quantum systems which are described by an entangled
wavefunction, the correlation of the corresponding components originates
from the state preparation of \textit{the primary source}. In our proposed
experiment, however, the correlation between the statistical values of the
spin components at t$_{1}$ and t$_{2}$ is a result of \textit{the past
history} of the particle which is due to the preparation of a new state at t$%
_{1}$ (denoted by $\varphi _{A}^{(t_{1})}$).

If we regard the violations of the Bell inequality for any quantum system
(including the case of two particles or the case of one particle) as a
consequence of the dependence of the state of the system on the preparation
conditions at a hidden variables level, the interpretation of Bell's theorem
on a unique basis would be possible. Our work demonstrates the significance
of such an interpretation. \strut \bigskip

{\LARGE Appendix \bigskip }\newline
\nolinebreak According to the statistical independence condition, which
leads to the relations (11) and (12) at the hidden variables level, one can
see how it is possible to reproduce the quantum predictions.

To begin with, there are some elementary relations which are valid as in the
two particle case, and they are given as follows\ \medskip

\begin{equation}
P^{(t_{1},t_{2})}(A,B|\widehat{a},\widehat{b},\Psi _{0})=\int_{\Lambda }\wp
^{(t_{1})}(A|\widehat{a},\lambda )\wp ^{(t_{2})}(B|\widehat{b},\lambda )\
\rho (\lambda )\ d\lambda  \tag{A-1}
\end{equation}

\begin{equation}
P^{(t_{1})}(A|\widehat{a},\Psi _{0})=\int_{\Lambda }\wp ^{(t_{1})}(A|%
\widehat{a},\lambda )\ \rho (\lambda )\ d\lambda  \tag{A-2}
\end{equation}
\begin{equation}
P^{(t_{2})}(B|\widehat{a},\widehat{b},A,\Psi _{0})=\dfrac{1}{P^{(t_{1})}(A|%
\widehat{a},\Psi _{0})}\times \int_{\Lambda }\wp ^{(t_{1})}(A|\widehat{a}%
,\lambda )\wp ^{(t_{2})}(B|\widehat{b},\lambda )\ \rho (\lambda )\ d\lambda 
\tag{A-3}
\end{equation}

\begin{equation}
\langle \sigma _{a}^{(t_{1})}\sigma _{b}^{(t_{2})}\rangle =\int_{\Lambda
}E^{(t_{1})}(\widehat{a},\lambda )\ E^{(t_{2})}(\widehat{b},\lambda )\ \rho
(\lambda )\ d\lambda  \tag{A-4}
\end{equation}
and

\begin{equation}
\langle \sigma _{a}^{(t_{1})}\rangle =\int_{\Lambda }E^{(t_{1})}(\widehat{a}%
,\lambda )\ \rho (\lambda )\ d\lambda  \tag{A-5}
\end{equation}
where $\langle \sigma _{a}^{(t_{1})}\rangle $ is the expectation value of
the spin component of the particle along $\widehat{a}$ at $t_{1}$. The
correspondence relations for $P^{(t_{2})}(B|\widehat{a},\widehat{b},\Psi
_{0})$ and $\langle \sigma _{b}^{(t_{2})}\rangle _{a}$ (the expectation
value of the spin component along $\widehat{b}$ at $t_{2}$, when the
particle has been already passed through \textbf{M}$_{1}(\widehat{a})$ at $%
t_{1} $), however, can be defined in a different way. To show this, we use
the following relations which hold for any dichotomic observables (here, $%
\sigma _{a}^{(t_{1})}$ and $\sigma _{b}^{(t_{2})}$), taking the values $\pm
1 $,

\begin{equation}
P^{(t_{2})}(B|\widehat{a},\widehat{b},\Psi _{0})=\dfrac{1}{2}(1+B\ \langle
\sigma _{b}^{(t_{2})}\rangle _{a})  \tag{A-6}
\end{equation}
and

\begin{eqnarray}
P^{(t_{2})}(B|\widehat{a},\widehat{b},A,\Psi _{0}) &=&\dfrac{%
P^{(t_{1},t_{2})}(A,B|\widehat{a},\widehat{b},\Psi _{0})}{P^{(t_{1})}(A|%
\widehat{a},\Psi _{0})}  \nonumber \\
&=&\dfrac{1}{2}\left[ 1+\frac{B\ \langle \sigma _{b}^{(t_{2})}\rangle
_{a}+AB\ \langle \sigma _{a}^{(t_{1})}\sigma _{b}^{(t_{2})}\rangle }{1+A\
\langle \sigma _{a}^{(t_{1})}\rangle }\right]  \tag{A-7}
\end{eqnarray}

According to relation (5), the conditional probabilities at $t_{2}$ do not
change, if we exchange the values $A$ and $B$. This means that

\begin{equation}
P^{(t_{2})}(B|\widehat{a},\widehat{b},A,\Psi _{0})=P^{(t_{2})}(A|\widehat{a},%
\widehat{b},B,\Psi _{0})  \tag{A-8}
\end{equation}
\medskip

If we impose the condition (A-8) on (A-7), we get

\begin{equation}
\frac{B\ \langle \sigma _{b}^{(t_{2})}\rangle _{a}+AB\ \langle \sigma
_{a}^{(t_{1})}\sigma _{b}^{(t_{2})}\rangle }{1+A\ \langle \sigma
_{a}^{(t_{1})}\rangle }=\frac{A\ \langle \sigma _{b}^{(t_{2})}\rangle
_{a}+AB\ \langle \sigma _{a}^{(t_{1})}\sigma _{b}^{(t_{2})}\rangle }{1+B\
\langle \sigma _{a}^{(t_{1})}\rangle }  \tag{A-9}
\end{equation}
which leads to a new relation

\begin{equation}
\langle \sigma _{b}^{(t_{2})}\rangle _{a}=\langle \sigma
_{a}^{(t_{1})}\rangle \ \langle \sigma _{a}^{(t_{1})}\sigma
_{b}^{(t_{2})}\rangle  \tag{A-10}
\end{equation}

Now, it is possible to define the correspondence relations for $%
P^{(t_{2})}(B|\widehat{a},\widehat{b},\Psi _{0})$ and $\langle \sigma
_{b}^{(t_{2})}\rangle _{a}$, using the relations (A-6), (A-10), (A-4) and
(A-5), as follows

\begin{equation}
\langle \sigma _{b}^{(t_{2})}\rangle _{a}=\int \int_{\Lambda }E^{(t_{1})}(%
\widehat{a},\lambda )\ E^{(t_{1},t_{2})}(\widehat{a},\widehat{b},\lambda
^{\prime })\ \rho (\lambda )\ \rho (\lambda ^{\prime })\ d\lambda \ d\lambda
^{\prime }  \tag{A-10}
\end{equation}

\begin{equation}
P^{(t_{2})}(B|\widehat{a},\widehat{b},\Psi _{0})=\dfrac{1}{2}\int
\int_{\Lambda }\left[ 1+E^{(t_{1})}(\widehat{a},\lambda )\ E^{(t_{1},t_{2})}(%
\widehat{a},\widehat{b},\lambda ^{\prime })\right] \rho (\lambda )\ \rho
(\lambda ^{\prime })\ d\lambda \ d\lambda ^{\prime }  \tag{A-11}
\end{equation}

\end{document}